\documentclass[sigconf]{acmart} 

\usepackage{booktabs} 
\usepackage[english]{babel}
\usepackage{subcaption}
\usepackage{mathrsfs}
\usepackage[inline]{enumitem}

\setcopyright{rightsretained}

\acmDOI{10.475/123_4}

\acmISBN{123-4567-24-567/08/06}

\copyrightyear{2017}
\acmYear{2017}
\setcopyright{rightsretained}
\acmConference{SIGIR 2017 Workshop on Neural Information Retrieval (Neu-IR'17)}{}{August 7--11, 2017, Shinjuku, Tokyo, Japan} \acmPrice{}
\def\ignore#1{}

\newenvironment{ppl}{\fontfamily{ppl}\selectfont}{}

\begin{document}
\title[ ]{Neural Matching Models for Question Retrieval and Next Question Prediction in Conversation}


 \author{Liu Yang$^1$ \quad Hamed Zamani$^1$ \quad Yongfeng Zhang$^1$ \quad Jiafeng Guo$^2$ \quad W. Bruce Croft$^1$}
 \affiliation{%
 	\institution{
 		$^1$ Center for Intelligent Information Retrieval, University of Massachusetts Amherst, Amherst, MA, USA \\
 		$^2$ CAS Key Lab of Network Data Science and Technology, Institute of Computing Technology, \\ Chinese Academy of Sciences, Beijing, China}
 }
 \email{{lyang, zamani, yongfeng, croft}@cs.umass.edu, guojiafeng@ict.ac.cn}

\begin{abstract}

The recent boom of AI has seen the emergence of many human-computer conversation systems such as Google Assistant, Microsoft Cortana, Amazon Echo and Apple Siri. We introduce and formalize the task of predicting questions in conversations, where the goal is to predict the new question that the user will ask, given the past conversational context. This task can be modeled as a ``sequence matching'' problem, where two sequences are given and the aim is to learn a model that maps any pair of sequences to a matching probability. Neural matching models, which adopt deep neural networks to learn sequence representations and matching scores, have attracted immense research interests of information retrieval and natural language processing communities. In this paper, we first study neural matching models for the question retrieval task that has been widely explored in the literature,  whereas the effectiveness of neural models for this task is relatively unstudied. We further evaluate the neural matching models in the next question prediction task in conversations. We have used the publicly available Quora data and Ubuntu chat logs in our experiments. Our evaluations investigate the potential of neural matching models with representation learning for question retrieval and next question prediction in conversations.  Experimental results show that neural matching models perform well for both tasks.
\end{abstract}

%
%


\keywords{\noindent Deep learning, neural conversational models, question retrieval, neural networks}

\settopmatter{printacmref=false, printfolios=false}

\maketitle

\section{Introduction}
\label{sec:intro}




\begin{table}[t]
	\centering
	\caption{Motivated examples of predicting questions in conversations and search. Ground truth labels are highlighted by different text colors, where blue means correct predictions and red means wrong predictions.}
	\label{table:motivatedExamples}
	\scriptsize
	\begin{ppl}
		\begin{tabular}{|p{8.5cm}|} 
			\hline
			\multicolumn{1}{|c|}{Predicting Questions in Conversations as in Ubuntu IRC Chat Rooms}\\
			\hline
			\underline{{\emph{Example 1}}}:\newline 
			Time: 2010-12-18 \\
			Conversation Context: \newline
			[17:23] $<$neohunter111$>$ Hello I have a problem with my mouse, is a microsoft wireless mouse 7000, when i press button6 or buttton 7 ubuntu recives a lot of press and realease events!! any ideas of how to solve this or how to search in google?? \newline 
			[17:24] $<$pksadiq$>$ neohunter111: does system $>$ preferences $>$ mouse has any option? \newline 
			[17:26] $<$neohunter111$>$ pksadiq yes the mouse works, the problem is that i set the boutton 6 and 7 (muse wheel to left o right) to change the desktop screen. and when i press it the desktop cube turns like crazy a lot of times, but before was working ok. \newline 
			[17:27] $<$pksadiq$>$ neohunter111: go to compiz settings in system $>$ preferences,a dn select 3D desktop plugin and change settings \newline 
			Predicted Question: \newline
			\textcolor{blue}{\textit{Where is 3d desktop plugin?}} (Correct) \newline
			\textcolor{red}{\textit{Is there a keyboard shortcut to change desktop?}} (Wrong) \newline
			\underline{{\emph{Example 2}}}:\newline 
			Time: 2011-12-22 \\
			Conversation Context: \newline 
			[15:59] $<$gplikespie$>$ Hello, I am new to Linux and am not sure how to move files from windows to linux, can anyone help? \newline 
			[16:02] $<$etroshica$>$ gplikespie, there is a variety of methods, depending on the file size and how much you want to learn. You can use some basic tools like gmail to Dropbox to send files. If it's a VM you can use shared directories. You can also set up a samba share. If you have ssh access, I recommend winscp, definitely one of the easiest tools to use. \newline 															
			Predicted Question: \newline
			\textcolor{blue}{\textit{VM is virtual machine, right?}} (Correct)\newline
			\textcolor{red}{\textit{Would there be any reason why I should use a 32 bit version of Ubuntu instead of 64 bit for a VM?}} (Wrong)
			\\
			\hline
			\hline
			\multicolumn{1}{|c|}{Predicting Questions in Search (Question Retrieval) as in Quora}\\
			\hline
			\underline{{\emph{Example 1}}}:\newline 
			Query Question: How can I learn Deep Learning quickly? \newline 
			Predicted Questions: \\
			\textcolor{blue}{\textit{What are the best resources to learn about deep learning?}}  (Correct)\\
			\textcolor{blue}{\textit{How do I learn deep learning in 2 months ?}} (Correct)\\
			\textcolor{red}{\textit{How is deep learning used in search engines?}} (Wrong) \newline
			\underline{{\emph{Example 2}}}:\newline 
			Query Question: What made Steve Jobs a great presenter?  \newline 
			Predicted Questions: \\
			\textcolor{blue}{\textit{How can I make a presentation attractively just like Steve Jobs?}} (Correct)\\
			\textcolor{blue}{\textit{What are the secrets behind Steve Jobs' excellent live product presentations?}} (Correct) \\
			\textcolor{red}{\textit{What was it like to deliver a presentation to Steve Jobs?}} (Wrong)\\
			\hline
		\end{tabular}
	\end{ppl}
	\vspace{-0.0cm}
\end{table}

Due to the ability of neural network models to go beyond term matching similarities as well as omitting the feature engineering steps, neural matching models have recently achieved state-of-the-art performance in a number of information retrieval tasks. However, the generality of these models to be applied on different tasks is relatively unstudied.

In this paper, we focus on two question ranking tasks. The first one is \emph{question retrieval}: retrieving similar questions in response to a specific question. This task is useful in question answering and community question answering (CQA) applications. For instance, finding similar questions could help to improve the question answering accuracy or can help to avoid asking duplicate questions in CQA websites. Although neural approaches have been widely applied to answer sentence selection \cite{DBLP:conf/naacl/HeL16, Severyn:2015:LRS:2766462.2767738, DBLP:conf/acl/TanSXZ16} and similar question identification \cite{DBLP:journals/corr/WangHF17}, the effectiveness of deep learning architectures for question retrieval is relatively unstudied. Therefore, we study a set of neural networks that can retrieve similar questions to a given question. 

The second task is relevant to conversation models. Building intelligent systems that could perform meaningful conversations with humans has been one of the long term goals of artificial intelligence. Human-computer conversation plays a critical role in many popular mobile search systems, intelligent assistants, and chat bot systems such as Google Assistant, Microsoft Cortana, Amazon Echo, and Apple Siri. Traditional conversational systems are based on hand designed logics and features with natural language templates, which usually only works for restricted and predictable conversational inputs \cite{Nakano:2000:WTB:1117736.1117753,Lemon:2006:IDS:1608974.1608986, DBLP:journals/pieee/YoungGTW13}. With rich big data resources on the Web, enhanced GPU computational infrastructures, and large amount of labels derived from crowd sourcing and online user behaviors, end-to-end deep learning methods have begun to show promising results on conversation response ranking and generation tasks \cite{DBLP:journals/corr/ShangLL15,DBLP:conf/sigir/YanSW16,DBLP:journals/corr/BordesW16,DBLP:conf/acl/LiGBSGD16,DBLP:conf/emnlp/LiMRJGG16,DBLP:journals/corr/SordoniGABJMNGD15}. According to these motivations, we focus on a new type of conversational response ranking problem as the second task in the paper: \emph{predicting the next question in a conversation}. During real conversations, humans could not only generate reasonable responses, but also have the ability to predict what the new questions that other speakers will be likely to ask. Learning models that could predict questions in conversations could enable us to better understand user intents during the conversations. Proactive content recommendations could be made without implicit questions issued by users. Furthermore,  pre-selected answer sets could be generated based on question prediction results as a cache mechanism to improve the efficiency and effectiveness of conversational question answering systems. Table  \ref{table:motivatedExamples} shows a number of motivated examples of predicting questions in conversations.

Our neural network architecture for both tasks is inspired by previous work \cite{DBLP:conf/naacl/HeL16, Severyn:2015:LRS:2766462.2767738, DBLP:conf/acl/TanSXZ16, DBLP:journals/corr/WangJ16b,DBLP:conf/sigir/YanSW16} that achieves impressive performance in different tasks. The designed siamese neural network models the long dependency of terms using a long short term memory (LSTM) layer. It further takes advantage of multiple convolutional and max pooling layers for representation learning of sequences based on the output of the LSTM layer. The network outputs a real-valued score for each candidate question and all candidate questions are ranked based on their matching score computed by the network.

We evaluate our models for the question retrieval task using the recently released Quora dataset. Our experiments demonstrate that the proposed neural network model outperforms state-of-the-art non-neural question retrieval approaches. The experiments also validate the hypothesis that neural matching models can complement exact term matching approaches in the question retrieval task; hence, a combination of the two is more appropriate. For the next question prediction task, we trained our model on the chat logs extracted from Ubuntu-related chat rooms on the Freenode Internet Relay Chat (IRC) network\footnote{\url{http://dataset.cs.mcgill.ca/ubuntu-corpus-1.0/}}. Our experiments suggest that neural matching models could perform well for both tasks, which demonstrates the potential of neural matching models with representation learning for new applications and scenarios.


The contributions of this work are as follows: (1) We introduce and formalize the new task of next question prediction in conversations. (2) We study of the effectiveness of neural matching models for question retrieval and predicting questions in conversations. Experimental results show that neural matching models perform well for both tasks. 

\section{Neural Matching Model}
\label{sec:deep_neural_match_model}

\subsection{Problem Definition and Model Overview}
\label{sec:model_overview}
In this section, we formally explain the high-level architecture of our model. Let matrices $\mathbf{Q} \in \mathbb{R}^{l \times |Q|}$ and $\mathbf{P} \in \mathbb{R}^{l \times |P|}$ denote the word embeddings of two sequences $\mathbf{Q}$ and $\mathbf{P}$, respectively. Let $|\cdot|$ denote the sequence length and $l$ denote the embedding dimensionality of individual vocabulary terms. Each column of $\mathbf{Q}$ and $\mathbf{P}$ is a word embedding vector representing a word in the sequences. Each sequence pair $\mathbf{Q}$ and $\mathbf{P}$ is associated with a label $y$. Given a query sequence $\mathbf{Q}$ and multiple candidate sequences $\{\mathbf{P}_1, \mathbf{P}_2 \cdots \mathbf{P}_n\}$, the goal is to generate a candidate sequence rank list. For the question retrieval task, the query sequence $\mathbf{Q}$ is a question and the candidate sequences $\mathbf{P}$ are the candidate similar questions. For the question prediction task in conversations, the query sequence $\mathbf{Q}$ is the previous context, consisting of previous questions with their responses, and the candidate sequences are the next candidate questions.

\begin{figure}[th]
	\center
	\includegraphics*[viewport=0mm 0mm 200mm 93mm, scale=0.4]{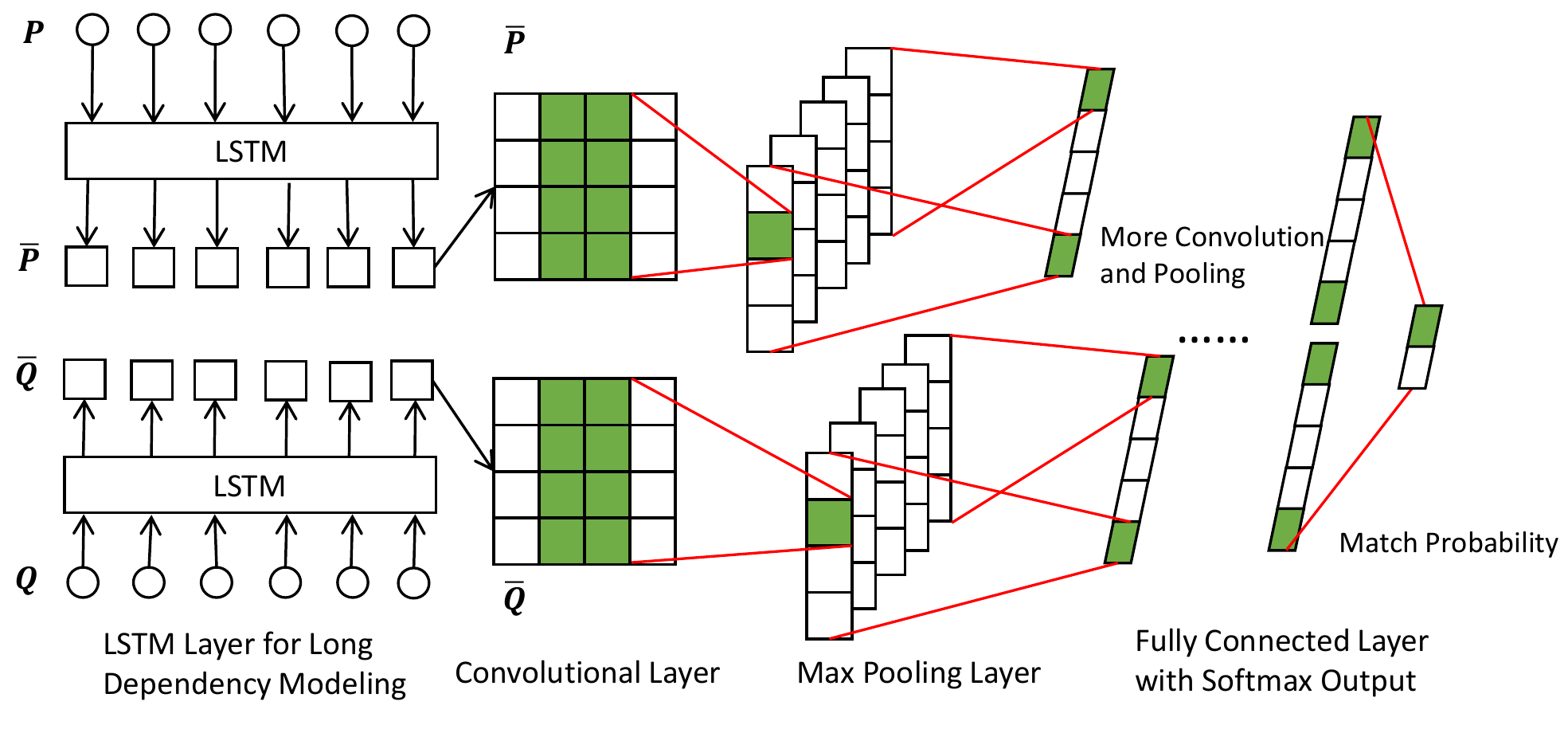}
	\caption{The architecture of LSTM-CNN-Match model for matching sequences.}\label{fig:blstm-cnn-match}
\end{figure}
Figure \ref{fig:blstm-cnn-match} shows the architecture of LSTM-CNN-Match model for matching sequences. This model is an extension of the CDNN model proposed by Severyn and Moschitti \cite{Severyn:2015:LRS:2766462.2767738} that has been also explored in various applications such as answer sentence selection \cite{DBLP:conf/naacl/HeL16, DBLP:conf/acl/TanSXZ16, DBLP:journals/corr/WangJ16b, DBLP:journals/corr/ZhouQZXBX16,DBLP:conf/sigir/YanSW16,DBLP:conf/cikm/YangAGC16}. Comparing with CDNN, this model adopts a long short term memory (LSTM) layer for long term dependency modeling in sequences. The convolutional layers are running on the output of the latent representations modeled by the LSTM layer, instead of the raw word embeddings sequence. In the following, we describe the model in more detail.



\subsection{LSTM for Long Term Dependency Modeling}
We use an LSTM layer to process $\mathbf{Q}$ and $\mathbf{P}$ for modeling long term dependency information in the sentences. LSTM \cite{Hochreiter:1997:LSM:1246443.1246450} is an advanced variant of recurrent neural networks (RNN). It can overcome the vanishing / exploding gradient problem of simpler Vanilla RNNs with the memory cell and gating mechanisms. Each LSTM cell consists of a memory cell that stores information over a long history and three gates that specify how to control the information flow into and out of the memory cell. Given an input sequence $\mathbf{Q} = (x_0, x_1, ..., x_t)$, where $x_t$ denotes the word embedding at position $t$, LSTM outputs a new representation matrix $\bar{\mathbf{Q}}$ that captures contextual information seen before in addition to the word at position $t$ itself based on the equations below:

\begin{eqnarray}
i_t &=& \delta(W^i x_t + U^i h_{t-1} + b^i) \\
f_t &=& \delta(W^f x_t + U^f h_{t-1} + b^f) \\
o_t &=& \delta(W^o x_t + U^o h_{t-1} + b^o) \\
u_t &=& \tanh(W^u x_t + U^u h_{t-1} + b^u)\\
c_t &=& i_t u_t + f_t c_{t-1}\\
h_t &=& o_t \tanh(c_t)\\ \nonumber
\end{eqnarray}
where $i,f,o$ denote the input, forget and output gates, respectively. $c$ is the stored information in the memory cells and $h$ is the learned representation. Thus $h_t$ is corresponding to the $t$-th column of the new representation matrix $\bar{\mathbf{Q}}$ which encodes the $t$-th word in $\mathbf{Q}$ with its context information. We also tried to use the bidirectional LSTM (Bi-LSTM). But we found that Bi-LSTM does not improve the performance. It led to lower training efficiency comparing with LSTM. Thus we just use one directional LSTM in our model.



\subsection{Convolutional and Max Pooling Layers}
Given the hidden representations learned by the LSTM layer, we use convolutional layers with different filter sizes and max pooling layers with different window sizes to learn sequence representations for generating the matching score. The convolution operation transforms the original feature map to a new feature map by moving the filters and computing the dot products of the filters with the corresponding feature map patch. Each filter slides over the whole embedding vectors, but varies in how many words it covers.\footnote{We set filter sizes to $[1, 2, 3, 4, 5]$ and use $128$ filters of each size in our model.} We slide the filters without padding the edges and perform a narrow convolution \cite{DBLP:journals/corr/KalchbrennerGB14} . We further feed the output of the convolutional layer to a rectified linear unit (ReLU) function which is simply defined as $\max(0,\mathbf{x})$ to add non-linearity. After that we apply a max pooling layer on the output of the ReLU function. Finally we use a fully connected layer with a softmax function to output the probability distribution over different labels.  

%

\subsection{Loss Function and Training}
%

We consider a pairwise learning setting during model training process. The training data consists of triples $(\mathbf{Q}_i, \mathbf{P}_i^+, \mathbf{P}_i^-)$ where $\mathbf{P}_i^+$ and $\mathbf{P}_i^-$ respectively denote the positive and the negative candidate sequence for $\mathbf{Q}_i$. The pairwise ranking-based hinge loss function is defined as:

\begin{eqnarray}
\mathcal{L} = \sum_{i=1}^{M} \max(0, \epsilon - S(\mathbf{Q}_i, \mathbf{P}_i^+) +  S(\mathbf{Q}_i, \mathbf{P}_i^-)) + \lambda ||\theta||^2_2
\end{eqnarray}
where $M$ is the number of triples in the training data. $\lambda ||\theta||^2_2$ is the regularization term where $\lambda$ and $\theta$ respectively denote the regularization coefficient and the model parameters. $\epsilon$ denotes the margin in the hinge loss. $S(\cdot, \cdot)$ denotes the output matching score from the last layer of the LSTM-CNN-Match model. The parameters of the network are optimized using the \textit{Adam} algorithm~\cite{DBLP:journals/corr/KingmaB14}. 

\section{Experiments}
\label{sec:exp}
\subsection{Datasets and Experimental Setting}
 

We use the publicly available datasets from Quora\footnote{\url{https://data.quora.com/First-Quora-Dataset-Release-Question-Pairs}} and Ubuntu IRC chat logs\footnote{\url{http://dataset.cs.mcgill.ca/ubuntu-corpus-1.0/}} for the experiments. The Quora dataset consists of $404,340$ lines of question pairs. Each line contains the IDs for each question in the pair, the full text for each question, and a binary value that indicates whether the line contains a similar question pair or not. To use this dataset for question retrieval evaluation, we conducted data sampling and pre-processing. There are $148,487$ similar question pairs in the Quora data, which form the positive question pairs. For each positive question pair, we randomly picked one of them as the query question $\mathbf{Q}$. Then the other question is the positive candidate question $\mathbf{P}^+$ for $\mathbf{Q}$.  We used negative sampling to construct the negative pairs following previous work \cite{DBLP:conf/aaai/WanLGXPC16}. Specifically for each query question $\mathbf{Q}$, we first used it to retrieve the top $1000$ results from the whole question set using Lucene\footnote{\url{http://lucene.apache.org/}} with BM25. Then we randomly selected $4$ questions from them except the known positive candidate question $\mathbf{P}^+$ to construct the negative candidate questions. Finally, we randomly separated the whole dataset to training, development and testing data with proportion $8:1:1$. The statistics of different data partitions of the Quora data is presented in Table \ref{tab:quora_data}.\footnote{Note that in some rare cases, the hits count for a query question returned by Lucene could be less than $4$. In this case, the actual candidate question number for this query question could be less than $5$.} 

For the Ubuntu chat log data, we also perform similar data sampling and pre-processing. We identify questions from dialogs by question marks. For each question $q^*$ in a dialog, we stochastically sample a pre-context size $c\in[2, C]$, where $C$ is the max number of questions in the pre-context.\footnote{In our experiments, we empirically set $C=6$. We skip a question if there are less than $2$ previous questions or the question length is less than $3$. We remove speaker IDs in candidate questions to insure that different methods rank questions by matching actual question content instead of spearker IDs. Words appear less than or equal to $5$ times are replaced by $<$UNK$>$.} Let $c' = \min(c,t)$, where $t$ is the total number of questions before $q^*$. Then we generate context for $q^*$ by merging previous $c'$ questions $\{q_1, q_2, \cdots, q_{c'}\}$ with their responses. Thus the true question response $q^*$ is the positive question candidate. We additionally randomly sample another $9$ negative question responses except the known positive candidate question following previous work \cite{DBLP:journals/corr/LowePSP15}. Finally, we randomly separated the whole dataset to training, development and testing data with proportion $8:1:1$. The statistics of different data partitions of the Ubuntu chat log data is presented in Table \ref{tab:ubuntu_data}.


For data pre-processing, we performed tokenization and punctuation removal. We maintained stopwords for neural models and removed them for the traditional retrieval models such as BM25 and QL. We used TensorFlow\footnote{\url{https://www.tensorflow.org/}} for the implementation of the neural matching models. 

 \begin{table}[t]
 	\centering
 	\caption{The statistics of Quora data.}
 	\begin{tabular}{ l || l | l  | l | l } 
 		\hline \hline
 		Data & Train & Dev & Test & Total \\\hline
 		\#QueryQ & 118,789 & 14,848 & 14,850 & 148,487 \\
 		\#CandidateQ & 593,932 & 74,240 & 74,250 & 742,422 \\
 		AvgQueryQLen & 9.81 & 9.88 & 9.87 & 9.85 \\
 		AvgCandidateQLen & 9.91 & 9.89 & 9.92 & 9.91 \\\hline \hline
 	\end{tabular}
 	\label{tab:quora_data}
 \end{table}
 
  \begin{table}[t]
  	\centering
  	\caption{The statistics of Ubuntu chat log data.}
  	\begin{tabular}{ l || l | l  | l | l } 
  		\hline \hline
  		Data & Train & Dev & Test & Total \\ \hline
	  	\#Context & 102,680 & 12,994 & 12,896 & 128,570 \\
	  	\#CandidateQ & 1,026,800 & 129,940 & 128,960 & 1,285,700 \\
	  	AvgContextLen & 125.85 & 125.40 & 125.44 & 125.76 \\
	  	AvgCandidateQLen & 14.59 & 14.56 & 14.55 & 14.59 \\ \hline \hline
  	\end{tabular}
  	\label{tab:ubuntu_data}
  \end{table}

\textbf{Word Embeddings.} We use Glove \cite{pennington2014glove} word embeddings, which are 300-dimension word vectors trained with a crawled large corpus with 840 billion tokens. Embeddings for words not present are randomly initialized with sampled numbers from a uniform distribution U[-0.25,0.25], which follows the same setting as in \cite{Severyn:2015:LRS:2766462.2767738}. 

\textbf{Additional Word Overlap Features.} As noted in previous work \cite{Severyn:2015:LRS:2766462.2767738, DBLP:journals/corr/YuHBP14}, one weakness of models relying on distributional word embeddings is their inability to deal with cardinal numbers and proper nouns. This also has impacts on matching question pairs or contexts with questions. Suppose we have two questions ``\textit{What happened in US in 1776?}'' and ``\textit{What happened in Japan in 1871?}''. These two questions will  be likely predicted with a high matching probabilities by neural matching models replying on word embedding input since country names like ``US'' and ``Japan'', numbers like ``1776'' and ``1871'' have close distances in the word embedding space. However, these two questions represent two different question intents. To mitigate this issue, we follow the approach in \cite{Severyn:2015:LRS:2766462.2767738, DBLP:journals/corr/YuHBP14} and include additional word overlap features into the model. Specifically, we compute the word co-occurrence count and IDF weighted word co-occurrence between two sequences. Computing these simple word overlap features is straightforward. We combine the matching probability learned by neural matching models with these two simple word overlap features with a logistic regression layer to generate the final ranking scores of candidate questions. 

\textbf{Model Hyper-parameters.}  We tuned the hyper-parameters with grid search using the development set. For the setting of LSTM-CNN-Match model in question retrieval, we set learning rate to $0.002$, batch size to $500$, margin of the hinge loss to $0.5$, filter sizes to $[1, 2, 3, 4, 5]$, and the number of each feature size to $128$. For the setting of LSTM-CNN-Match model in question prediction in conversations, we set learning rate to $0.002$, batch size to $200$, margin in the hinge loss to $0.3$, filter sizes to $[1, 2, 3, 4, 5]$, and the number of each feature size to $128$. 


\subsection{Evaluation Metrics and Compared Methods}
For the Quora data and Ubuntu chat log data, since there is only one positive candidate question for each query question or previous conversation context, we adopt mean reciprocal rank (MRR) and precision at the highest position (P@1) as the evaluation metrics. Note that in this case MRR is equivalent to MAP and P@1 is equivalent to R-Precision. For Ubuntu chat log data, since there are $10$ candidate questions for each context, we additionally report P@5 and Recall@5. We study the effectiveness of the following  methods:

\textbf{WordCount}: This method computes the word co-occurrence count between the two sequences.

\textbf{WordCountIDF}: This method computes the word co-occurrence count weighted by IDF value between the two sequences.

\textbf{VSM}: This method computes the cosine similarity between the TF-IDF representation of the given two sequences.

\textbf{BM25}: This method computes the BM25 score between the two sequences, where we treat one of the sequences as the query and the other one as the document.

\textbf{QL}: This method computes the query likelihood \cite{Ponte:1998:LMA:290941.291008} score with Dirichlet prior smoothing between the language models of the two sequences.

\textbf{TRLM}: This method is the translation-based language model employed by Jeon et al. \cite{Jeon:2005:FSQ:1099554.1099572} and Xue et al. \cite{Xue:2008:RMQ:1390334.1390416}. This method has been consistently reported as the state-of-the-art method for the question retrieval task.\cite{Zhou:2013:TFB:2505515.2505550}.

\textbf{AvgWordEmbed}: This method uses the average vector of word embeddings as the sequence representation; then the cosine similarity of sequence representations is used for the candidate question ranking.

\textbf{CNN-Match}: This is a degenerate version of the LSTM-CNN-Match model where we remove the LSTM layer in the model, which is similar to the CDNN model proposed by Severyn and Moschitti~\cite{Severyn:2015:LRS:2766462.2767738}. 
	
	
\textbf{LSTM-CNN-Match}: The model presented in Section \ref{sec:deep_neural_match_model}, which has been recently applied to other tasks, such as answer sentence selection \cite{DBLP:conf/naacl/HeL16, DBLP:conf/acl/TanSXZ16, DBLP:journals/corr/WangJ16b, DBLP:journals/corr/ZhouQZXBX16,DBLP:conf/sigir/YanSW16}.

\textbf{Combined Model}: We tried to combine scores of all baseline methods with neural matching models and trained a LambdaMART ranker for question ranking. This is to study whether combining learned features from basic retrieval models with neural models could lead to better retrieval performance.

\subsection{Experimental Results on Question Retrieval}

 \begin{table}[t]
 	\centering
 	\caption{Experimental results for question retrieval with the Quora dataset. The best performance is highlighted in boldface. $\ddagger$ means significant difference over all the baseline methods with $p < 0.05$ measured by the Student's paired t-test.}
 	\begin{tabular}{ l || l | l  } 
 		\hline \hline
 		Method & MRR & P@1 \\\hline
 		WordCount & 0.786 & 0.659 \\
 		WordCountIDF & 0.811 & 0.699 \\
 		VSM & 0.833 & 0.737 \\
 		BM25 & 0.861 & 0.781 \\
 		QL & 0.859 & 0.777 \\
 		TRLM  & 0.865 & 0.778 \\
 		AvgWordEmbed & 0.791 & 0.669 \\
 		CNN-Match  & 0.864 & 0.774 \\
 		LSTM-CNN-Match & 0.880 & 0.797 \\
 		Combined Model & \textbf{0.894}$^\ddagger$ & \textbf{0.819}$^\ddagger$ \\ \hline\hline
 	\end{tabular}
 	\label{tab:quora_res}
 \end{table}

Table \ref{tab:quora_res} shows the experimental results for the question retrieval task with the Quora dataset. We summarize our observations as follows: (1) LSTM-CNN-Match model outperforms all the baseline methods including basic retrieval models, translation model based methods and basic neural model/word embedding based methods. This shows the advantage of jointly modeling semantic match information through a neural matching model and basic word overlap information for the question retrieval task. (2) Comparing the performance of LSTM-CNN-Match model and CNN-Match model, we found that the retrieval performance will decrease if we remove the LSTM layer. This shows that modeling long term dependency in questions through LSTM is useful for boosting question search performance. (3) If we combine the learned matching score of neural models with the basic retrieval model scores, we can observe further gain over the baselines. Thus in practice the learning to rank framework is still useful for combining different features including both traditional IR model scores and the more recent neural model scores for a strong ranker for question search. 

%
To get a better understanding of the effectiveness of the model, we checked the retrieved questions of each method. Jointly modeling term matching information with semantic matching information is important for the question retrieval task. Table \ref{tab:ret_examples} reports the retrieval results of different methods for the query question ``What are some good anime movies?''. BM25 relying on term matching between question pairs ranked the correct similar candidate question ``What are some of the best anime shows?'' in a relatively low position and ranked ``What are good scary movies?'' in the first position. TRLM suffers from a similar problem. The neural matching model LSTM-CNN-Match ranked the correct similar question candidate in the first position, since it can capture the semantic similarity between ``movies''  and ``shows'' as well as ``good'' and ``best'', which are missed by the term matching based retrieval models.

 \begin{table}[t]
 	\centering
 	\caption{Retrieval results for the query question ``What are some good anime movies?'' of different methods. The correct similar candidate question is highlighted in bold font.}
 	\begin{tabular}{ l || l  } 
 		\hline \hline
 		Top  Retrieval Results by BM25 & Rank \\\hline
 		What are good scary movies ? & 1 \\
 		\textbf{What are some of the best anime shows?} & 4 \\\hline\hline
 		Top  Retrieval Results by TRLM & \\\hline 
 		What are good scary movies ? & 1 \\
 		\textbf{What are some of the best anime shows?} & 2 \\\hline\hline
 		Top  Retrieval Results by LSTM-CNN-Match &  \\\hline
 		\textbf{What are some of the best anime shows?} & 1 \\
 		What is your favorite anime ? & 2 \\\hline\hline
 	\end{tabular}                                              
 	\label{tab:ret_examples}
 \end{table}
 
  \begin{table}[t]
  	\centering
  	\caption{Experimental results for predicting questions in conversations with the Ubuntu chat log dataset. The best performance is highlighted in boldface. R@5 denotes Recall@5.}
  	\begin{tabular}{ l || l | l | l | l } 
  		\hline \hline
  		Method & MRR & P@1 & P@5 & R@5  \\\hline
  		WordCount & 0.474 & 0.284 & 0.143 & 0.717 \\
  		WordCountIDF & 0.548 & 0.391 & 0.146 & 0.732 \\
  		VSM & 0.570 & 0.432 & 0.146 & 0.729 \\
  		BM25 & 0.559 & 0.413 & 0.146 & 0.728 \\
  		QL & 0.483 & 0.337 & 0.127 & 0.633 \\
  		CNN-Match & 0.579 & 0.428 & \textbf{0.155} & \textbf{0.775} \\
  		LSTM-CNN-Match & 0.571 & 0.426 & 0.151 & 0.754 \\
  		Combined Model & \textbf{0.581} & \textbf{0.440} & 0.152 & 0.762 \\ \hline\hline
  	\end{tabular}
  	\label{tab:ubuntu_res}
  \end{table}
 
 \subsection{Experimental Results on Predicting Questions in Conversations}
 
Table \ref{tab:ubuntu_res} shows the experimental results for predicting questions in conversations with the Ubuntu chat log dataset.  For this task, the ``Combined Model'' performed the best for MRR and P@1. CNN-Match achieved the best performances for P@5 and Recall@5. We also found LSTM-CNN-Match performed worse than CNN-Match for this task. Overall neural matching models could improve the ranking effectiveness of finding questions given previous context over traditional retrieval models. Combining scores from neural matching models and traditional retrieval models could also be helpful. Our research represents an initial effort to understand the effectiveness of neural matching models for predicting questions in conversations. We find that this is a more challenging task comparing with similar question finding due to at least two reasons: 1) Unlike similar question pairs with close sequence lengths, a context is usually much longer than a candidate question in conversations. 2) The matching pattern between conversational context and candidate questions could be more complex, which is beyond semantic match or paraphrase as in question retrieval. To find more effective clues from context, more advanced model architectures like attention modeling in context should be considered. Sequence to sequence learning with an RNN Encoder-Decoder architecture \cite{Sutskever:2014:SSL:2969033.2969173, DBLP:journals/corr/ChoMGBSB14, DBLP:journals/corr/ShangLL15} and memory networks \cite{Sukhbaatar:2015:EMN:2969442.2969512} could be promising directions to explore.

\section{Related Work}
\label{sec:rel}

\subsection{Question Retrieval}



The current research for question retrieval can be divided into two categories. The first group leveraged translation models to bridge the lexical gaps between questions. Jeon et al. \cite{Jeon:2005:FSQ:1099554.1099572} proposed a method learning word translation probabilities from question-question pairs collected based on similar answers in CQA. Xue et al. \cite{Xue:2008:RMQ:1390334.1390416} proposed a retrieval model that combines a translation-based language model for the question part with a query likelihood approach for the answer part. The translation-based language model (TRLM) has been consistently reported as the state-of-the-art method for question retrieval \cite{Zhou:2013:TFB:2505515.2505550}. Topic models have also been adopted for question retrieval \cite{Yang:2013:CJM:2505515.2505720}. Recent years there are few research works on the research of building deep learning models with word embeddings for question retrieval \cite{DBLP:conf/acl/ZhouHZH15,Wang:2017:CEC:3018661.3018687}. Wang et al.\cite{Wang:2017:CEC:3018661.3018687} proposed a unified framework to simultaneously handle the three problems in question retrieval including lexical gap, polysemy and word order A high level feature embedded convolutional semantic model is proposed to learn the question embeddings.

The second research group has focused on improving question search with category information about questions. Cao et al. \cite{Cao:2009:UCI:1645953.1645989} proposed a language model with leaf category smoothing for questions in the same category.  Zhou et al. \cite{Zhou:2013:TFB:2505515.2505550} proposed an efficient and effective retrieval model for question retrieval by leveraging user chosen categories. They achieved this by filtering some irrelevant historical questions under a range of leaf categories. Although considering category information can improve question retrieval performance, these methods could not be applied to the scenarios where the category information is not available. In many question answering and chatbot/dialogue systems, new questions issued by users have no explicit predefined category. Our work is closer to a general setting of question search where no category information are available.

%
%
\subsection{Neural Conversation Models}
Recent years there are growing interests on research about conversation response generation and ranking with deep learning and reinforcement learning \cite{DBLP:journals/corr/ShangLL15,DBLP:conf/sigir/YanSW16,DBLP:journals/corr/BordesW16,DBLP:conf/acl/LiGBSGD16,DBLP:conf/emnlp/LiMRJGG16,DBLP:journals/corr/SordoniGABJMNGD15}. Shang et al. \cite{DBLP:journals/corr/ShangLL15} proposed Neural Responding Machine
(NRM), which is a RNN encoder-decoder framework for short text conversation and showed that it outperformed retrieved-based methods and SMT-based methods for single round conversation. Sordoni et al. \cite{DBLP:journals/corr/SordoniGABJMNGD15} proposed a neural network architecture for response generation that is both context-sensitive and data-driven utilizing the Recurrent Neural Network Language Model architecture. Yan et al. \cite{DBLP:conf/sigir/YanSW16} proposed
a retrieval-based conversation system with the deep learning-to-respond
schema through a deep neural network framework driven
by web data. Li et al. \cite{DBLP:conf/emnlp/LiMRJGG16} apply deep reinforcement learning to model future reward in chatbot dialogs towards building a neural conversational model based on the long-term success of
dialogs. Bordes et al. \cite{DBLP:journals/corr/BordesW16} proposed a
testbed to break down the strengths and shortcomings of end-to-end dialog systems in goal-oriented applications. They showed that an end-to-end dialog system based on Memory Networks can reach promising performance and learn to perform non-trivial operations. We work is relevant to neural conversational models. But we have different focuses on finding questions given previous conversational context.

\subsection{Neural Ranking Models}
A number of neural approaches have been proposed for ranking documents in response to a given query.
These approaches can be generally divided into two groups: representation-focused and interaction-focused models~\cite{Guo:2016}. Representation-focused models independently learn a representation for each query and candidate document and then calculate the similarity between the two estimated representations via a similarity function. As an example, DSSM \cite{Huang:2013} is a feed forward neural network with a word hashing phase as the first layer to predict the click probability given a query string and a document title. 

On the other hand, the interaction-focused models are designed based on the interactions between the query and the candidate document. 
For instance, DeepMatch~\citep{Lu:2013} is an interaction-focused model that maps each input to a sequence of terms and trains a feed-forward network to compute the matching score. These models have an opportunity to capture the interactions between query and document, while representation-focused models look at the inputs in isolation. Recently, Mitra et al.~\citep{Mitra:2017} proposed to simultaneously learn local and distributional representations to capture both exact term matching and semantic term matching.

All the aforementioned models are trained based on either explicit relevance judgments or clickthrough data. More recently, Dehghani et al. \cite{Dehghani:2017} proposed to train neural ranking models when no supervision signal is available. They used an existing retrieval model, e.g., BM25 or query likelihood, to generate large amount of training data automatically and proposed to use these generated data to train neural ranking models with weak supervision.
\section{Conclusions and Future Work}
\label{sec:conclu}
In this paper, we studied the effectiveness of neural matching models for two tasks: retrieving similar questions and predicting questions in conversations. We showed that neural matching models significantly outperforms all the baseline methods for the question retrieval task. Furthermore, when the neural matching model is combined with the basic term matching based retrieval models, we can achieve larger gains. For predicting questions in conversations, we observed that LSTM layers cannot handle long question history (past questions) and thus a simpler neural matching model with no LSTM layer outperforms all the other methods. This is a preliminary study in this area and there are still spaces to develop more advanced neural models to further improve the performance of matching conversational context with questions. For future work, we plan to continue the research on neural conversational models as a modern way for people to access information. Modeling context attentions and incorporating external knowledge into neural conversation models for finding better candidate questions could be also considered as interesting future directions.

\section{Acknowledgments}
This work was supported in part by the Center for Intelligent Information Retrieval, in part by NSF IIS-1160894, and in part by NSF grant \#IIS-1419693. Any opinions, findings and conclusions or recommendations expressed in this material are those of the authors and do not necessarily reflect those of the sponsor.
\bibliographystyle{ACM-Reference-Format}
 \bibliography{similarQRet}  
\end{document}